\documentclass[draft,11pt]{article}
\usepackage{amssymb}
\begin{document}

\newcommand{\eq}{\triangleq}
\newcommand{\x}{{\bf x}}
\newcommand{\y}{{\bf y}}
\newcommand{\z}{{\bf z}}
\newcommand{\w}{{\bf w}}
\renewcommand{\u}{{\l}}
\newcommand{\e}{{\bf e}}
\newcommand{\0}{{\bf 0}}
\newcommand{\varliminf}{\mathop{\underline{\lim}}\limits}
\newcommand{\varlimsup}{\mathop{\overline{\lim}}\limits}
\newcommand{\p}{{\sf p}}
\newcommand{\A}{{\sf A}}
\renewcommand{\l}{{\ell}}
\newcommand{\D}{{\cal D}}
\newcommand{\M}{{\cal M}}
\renewcommand{\S}{{\cal S}}
\newcommand{\f}{{\sf f}}

\def\NN{{\mathbb N}}
\def\QQ{{\mathbb Q}}
\def\RR{{\mathbb R}}
\def\ZZ{{\mathbb Z}}

\begin{center}
{\Large\bf
   Superimposed  Codes and Threshold
 Group Testing}\footnote[1]{An extended version of this draft was
 published in:  {\em Lecture Notes in Computer Science}, vol. 7777, pp. 509-533,
 2013.}
\\[15pt]
{\bf A. D'yachkov, \quad V.  Rykov, \quad C. Deppe, \quad V. Lebedev}
\\[15pt]
Moscow State University, Faculty of Mechanics and
Mathematics,\\
Department of Probability Theory, Moscow, 119992, Russia,\\
{\sf agd-msu@yandex.ru,\quad vrykov@unomaha.edu}
\end{center}

\textbf{Abstract-} {\sf We will discuss superimposed codes and non-adaptive
group testing designs arising from the potentialities of compressed
genotyping models in  molecular biology. The given paper was
motivated by the 30th anniversary of D'yachkov-Rykov  recurrent upper bound
on the rate of  superimposed codes published in~1982.
We were also inspired by recent results obtained
for non-adaptive threshold group testing
which   develop the theory of superimposed codes.}
\medskip

{\sl Index terms}. {\sf Group testing, compressed
genotyping, screening experiments, search designs, superimposed codes,
 rate of codes, rate of designs, bounds on the rate, shortened RC-code,
threshold search designs}.


\section{Introduction}
\quad
We consider superimposed codes and  non-adaptive group testing models.
These search models are also termed as
combinatorial  designs of screening  experiments or pooling designs.
Designing  screening experiments (DSE)~(\cite{Maljutov1},
\cite{Maljutov2}, \cite{Maljutov3}) can be located in applied
mathematics in the border region of search and information
theory~\cite{Renyi},\cite{such}.
In many ``processes'' which are dependent on a large number of factors, it is
natural, that one assumes a small number of ``significant'' factors, which really
control the process, and considers the influence of the other factors as mere
``experiment errors''. Experiments to identify the significant factors are
called screening experiments.

A typical problem from DSE theory called a symmetric model of DSE~\cite{d04}
or symmetric search model is the following.
Among $t$ factors there are $p$ ``significant'', which need to be identified.
By $N$ tests which  examine arbitrary distinct $N$ subsets of the factors,
it can be determined $N$ values of a function {\em depending only on the number of significant
factors included in the tests}. One tries to perform these
experiments as economical as possible. The main criterion  at this is the search
duration: how many tests $N$ are at least necessary to identify all significant
factors in the most unfavorable case?

The  aim of our paper is to present the principal combinatorial
results for the symmetric  search model. We don't  discuss here the
general noisy symmetric model of non-adaptive search designs which can be described
using the terminology of multiple access channel  (MAC)~\cite{ck81}.
An interested reader is referred to~\cite{d04}. The information theory problems  for non-symmetric
search model are considered in~\cite{Maljutov3}.

The paper is organized as follows.
In Section~2, we give a brief survey of necessary  definitions and bounds on the
rate of superimposed codes which are the {\em base for studying} of non-adaptive
group testing models.

In Section~3, we introduce the concept of
non-adaptive group testing designs arising from the potentialities of compressed
genotyping models in  molecular biology and establish a universal
upper bound on their rate. The universal bound is prescribed by
D'yachkov-Rykov~\cite{dr82} recurrent upper bound on the rate of classical superimposed
codes.

In Section~4, we remind  our constructions of superimposed codes
based on shortened Reed-Solomon codes (RS-codes)~\cite{dmr00_1}-\cite{dmtv02}
and other ideas~\cite{KL04}-\cite{lk04}.
In these  papers we essentially   extended  optimal
and suboptimal construction of classical  superimposed
codes suggested in~\cite{ks64}. Note that  we included
in~\cite{dmr00_1}-\cite{dmtv02}
the detailed tables with parameters of the best known superimposed
codes. We don't mention  other authors because, unfortunately, we
don't know any papers containing relevant  results, i.e.,
the similar or improved tables of parameters.
Any extension  of our tables is the important open problem.

In Section~5, the threshold group testing model is discussed.
We apply the conventional terminology of superimposed
code theory to refine the description of a new lower bound on the
rate of threshold designs recently obtained in~\cite{mahdi-11}.

\subsection{Notations, Definitions and Relevant Issues}
\quad
Let  $[n]$ be the set of integers from 1 to~$n$ and the symbol $\eq$
denote definitional equalities.
For integers $N\ge2$ and $t\ge2$, symbols
$\Omega_j\subset[N]$, $j=1,2,\dots,t$, denote subsets of $[N]$.
Subsets $\Omega_j$, $j\in[t]$,
are identified with  binary
columns $\x(j)\eq(x_1(j),x_2(j),\dots,x_N(j))$ in which
$$
x_i(j)\eq\cases{1 & if $i\in\Omega_j$,\cr
              0 & if $i\not\in\Omega_j$,\quad
$i\in[N]$.\cr}
$$
 An incidence matrix $X\eq\|x_i(j)\|,\;i\in[N],\;j\in[t]$,
is called a {\em code} with $t$ codewords
(columns) $\x(1),\x(2),\dots,\x(t)$ of length $N$
corresponding to a {\em family of subsets}
$\;\Omega_1,\,\Omega_2,\dots,\Omega_t$.

Let $P\subset[t]$ be an arbitrary fixed subset of $[t]$ and
$|P|$ is its size, i.e.,
$$
P\,\eq\,\left\{p_1,p_2,\dots,p_{|P|}\right\}\,\subset[t],
\quad 1\le p_1<p_2<\cdots<p_{|P|}\le t.
$$
  Denote by ${\cal P}(t,\le s)$ $\left({\cal P}(t,=s)\right)$ the
collection of all $\sum_{i=0}^s{t\choose i}$  $\left({t\choose s}\right)$
subsets $P$ of size $|P|\le s$ $\left(|P|=s\right)$.  Let
$N\ge2$ be an integer and
$\A=\{A_1,A_1,\dots,A_N\},\quad A_i\subset[t],\quad i\in[N],$
is a fixed family of subsets of $[t]$.
Subsets $A_i$ are identified with  binary
rows $\x_i\eq(x_i(1),x_i(2),\dots,x_i(t))$ in which
$$
x_i(j)\eq\cases{1 & if $j\in A_i$,\cr
              0 & if $j\not\in A_i$,\quad
$i\in[N],\;j\in[t].$\cr}
$$
We will identify the family $\A$ with its incidence
matrix (code)~$X=\|x_i(j)\|$, $\;i\in[N]$, $\;j\in[t]$.

In the theory of  {\em group testing}~\cite{dh93} ({\em designing screening
experiments}~\cite{d04})
the given, in advance, family $\A=\{A_1,A_1,\dots,A_N\}$ is interpreted as
a  {\em non-adaptive search design} consisting  of $N$ group tests
(experiments) $A_i$, $i\in[N]$. An experimenter wants
to construct group  tests  $A_i$, $i\in[N]$, to carry out the corresponding
experiments  and then
to identify an {\em unknown subset} $P\subset[t]$
with  the help of test outcomes  provided that
$P\subset{\cal P}(t,\le s)$ or~$P\subset{\cal P}(t,=s)$, where~$s\ll t$.
If for each  test $A_i$, $i\in[N]$, its outcome
{\em depends only on the size of intersection}
$$
|P\cap A_i|\,=\,\sum\limits_{m=1}^{|P|}\,x_i(p_m),\quad i\in[N],
$$
then we will say that  a {\em symmetric model}~\cite{d04}
of non-adaptive search design is considered.

\section{Superimposed $(z,u)$-Codes}
\quad
In this section we give a brief survey of necessary  definitions and bounds on the
rate of superimposed codes which are the {\em base for studying} of non-adaptive
group testing models.

Let $z$ and $u$ be positive integers such that~$z+u\le t$.

\textbf{Definition 1.} \cite{dmtv02}.  A family of subsets
$\Omega_1,\Omega_2,\dots,\Omega_t$, where
$\Omega_j\subseteq[N]$, $j\in[t]$,  is called
an $(z,u)$--{\em cover-free family} if for any two non-intersecting
 subsets $Z,\,U\subset[t]$, $Z\cap U=\emptyset$,
such that $|Z|=z$, $|U|=u$,
the following condition holds:
$$
\bigcap\limits_{j\in U}\,\Omega_j\, \not\subseteq
\bigcup\limits_{j\in Z}\,\Omega_j.
$$
 An incidence matrix
$X=\|x_i(j)\|,\;i\in[N],\;j\in[t]$,
corresponding to  $(z,u)$--cover-free
family is called a {\em superimposed $(z,u)$-code}.
\medskip

The following evident  necessary and
sufficient condition for Definition~1 takes place.
\smallskip

\textbf{Proposition 1.} \cite{dmtv02}. {\em Any binary $(N\times t)$-matrix
$X$ is a  superimposed $(z,u)$-code if and only if
for any two subsets $Z,\,U\subset[t]$,
such that $|Z|=z$, $|U|=u$ and $Z\cap U=\emptyset$
the matrix $X$ contains a row $\x_i=(\,x_i(1),x_i(2)\dots,x_i(t)$,
for which}
$$
x_i(j)=1\quad\mbox{for all}\quad j\in U,\qquad
x_i(j)=0\quad\mbox{for all}\quad j\in Z.
$$

Let  $\,t(N,z,u)$  be the maximal possible size
of superimposed  $\,(z,u)$-codes.
For fixed $1\le u<z$, define a {\em rate} of $\,(z,u)$-codes:
$$
R(z,u)\eq\varlimsup_{N\to\infty}
\frac{\log_2t(N,z,u)}{N}.
$$
For the classical case $u=1$, superimposed $(z,1)$--codes
and their applications were introduced by W.H~Kautz, R.C.~Singleton
in~\cite{ks64}.
Further, these codes along with new  applications were investigated
in~\cite{dr82}-\cite{d04}.
The best known upper and lower bounds on the rate $R(z,1)$ can be found 
in papers~\cite{dr82},\cite{drr89} and~\cite{dmtv02}.

\subsection{Recurrent Upper Bounds on $R(z,1)$ and $R(z,u)$}
\quad
Let $h(\alpha)\eq-\alpha\log_2\alpha-(1-\alpha)\log_2(1-\alpha)$,
$0<\alpha<1$, be the
binary entropy.  To formulate an {\em upper bound} on
the rate $R(z,1)$, $z\ge1$, we introduce the
function~\cite{dr82}
$$
\f_z(\alpha)\eq h(\alpha/z)-\alpha\, h(1/z),\quad z=1,2,\dots,
$$
of argument $\alpha$, $0<\alpha<1$.
\medskip

\textbf{Theorem 1.} \cite{dr82}-\cite{dr83}.\quad
(Recurrent upper bound on $R(z,1)$).
{\em If $z=1,2,\dots$, then the
rate $R(z,1)\le\overline{R}(z,1)$, where
$$
\overline{R}(1,1)=R(1,1)=1,\quad
\overline{R}(2,1)\eq\max_{0<\alpha<1}\;\f_2(\alpha)\,=\,0.321928
\eqno(1)
$$
and sequence $\overline{R}(z,1)$, $z=3,4,\dots$, is
defined  as the unique solution of  recurrent equation}
$$
\overline{R}(z,1)=\,\f_z\left(1-
\frac{\overline{R}(z,1)}{\overline{R}(z-1,1)}\right).
\eqno(2)
$$

Up to now, the recurrent
sequence $\overline{R}(z,1)$, $z=1,2,\dots$,
defined by (1)-(2) and called a {\em recurrent upper bound} has been the
best known upper bound on the rate~$R(z,1)$.
The reciprocal values of $\overline{R}(z,1)$, 
 $z=2,3,\dots,17$, taken from~\cite{dr83},
are given in Table~1.
\begin{center}
\begin{tabular}{|c|c||c|c||c|c||c|c|}
\hline
$z$ & $1/\overline{R}(z,1)$ & $z$ & $1/\overline{R}(z,1)$
& $z$ & $1/\overline{R}(z,1)$ & $z$ & $1/\overline{R}(z,1)$\\
\hline
\hline
$2$ & 3.1063 & 6 & 12.0482 & 10 & 24.5837 & 14 & 40.3950\\
\hline
$3$ & 5.0180 & 7 & 14.8578 & 11 & 28.2402 & 15 & 44.8306\\
\hline
$4$ & 7.1196 & 8 & 17.8876 & 12 & 32.0966 & 16 & 49.4536\\
\hline
$5$ & 9.4660 & 9 & 21.1313 & 13 & 36.1493 & 17 & 54.2612\\
\hline
\end{tabular}
\end{center}
\centerline{\textbf{Table 1.}}
\smallskip

Applying Theorem~1 and the corresponding calculus arguments, we
proved
\smallskip

\textbf{Theorem~2}. \cite{dr82}-\cite{dr83}.\quad
(Non-recurrent upper bound on $R(z,1)$).
{\em For any $z\ge2$, the rate $R(z,1)$
satisfies  inequality
$$
R(z,1)\,\le\,\frac{2\log_2[e(z+1)/2]}{z^2},\qquad z=2,3,\dots,
$$
which leads to the asymptotic inequality}
$$
R(z,1)\,\le\,\frac{2\log_2z}{z^2}\,(1+o(1)), \quad
z\to\infty.
$$
\medskip

\textbf{Theorem 3.}~\cite{l03}\quad
 (Recurrent inequality for $R(z,u)$).
{\em
If $z\ge\,u\,\ge2$, then for any $i\in[z-1]$ and
$j\in[u-1]$, the rate}
$$
R(z,u)\,\le\;
\frac{R(z-i,u-j)}{R(z-i,u-j)+\frac{(i+j)^{i+j}}{i^i\cdot j^j}}.
\eqno(3)
$$

Recurrent inequality (3) and the known numerical values of
recurrent upper bound  $\overline{R}(z,1)$, $z=1,2,\dots$,
defined by~(1)-(2), give  numerical values of
the best known  upper bound $\overline{R}(z,u)$ on
the rate~$R(z,u)$, $z\ge\,u\,\ge2$.
An asymptotic consequence from the given upper bound
is presented by
\medskip

\textbf{Theorem 4.}~\cite{dvy02}\quad
{\em If $z\to\infty$ and $u\ge2$ is fixed, then}
$$
R(z,u)\,\le\,\overline{R}(z,u)\,\le\,
\frac{(u+1)^{u+1}}{2\,e^{u-1}}\cdot
\frac{\log_2\,z}{z^{u+1}}\,\cdot(1+o(1)).
$$

\subsection{Random Coding Lower Bounds on $R(z,u)$ and $R(z,1)$}
\quad
\textbf{Theorem 5.}~\cite{dmtv02}\quad
{\em A random coding lower bound on the rate $R(z,u)$
has the form:
$$
R(z,u)\,\ge\,\underline{R}(z,u)\eq-(z+u-1)^{-1}
\log_2\left(1-\frac{z^z\,u^u}{(z+u)^{z+u}}\right),
\quad 2\le u< z.
$$
If $u\ge2$ is fixed and $z\to\infty$, then the asymptotic form of
the given lower bound is}
$$
R(z,u)\,\ge\,\underline{R}(z,u)=\frac{e^{-u}\cdot u^{u}\cdot\log_2e}{z^{u+1}}
\,\cdot(1+o(1)).
$$

If $u=1$, then the best known random coding lower bound on the rate $R(z,1)$
is given by
\smallskip

\textbf{Theorem 6.}~\cite{dr89}\quad  {\em For any $z=1,2,\dots$, the rate
$
R(z,1)\ge\underline{R}(z,1)\eq\frac{A(z)}{z},
$
 where
$$
A(z)\,\eq\,\max_{0<\alpha<1,\;0<Q<1}\,\left\{-(1-Q)\log
(1-\alpha^{z})+z\left(Q\log
\frac{\alpha}{Q}+(1-Q)\log\frac{1-\alpha}{1-Q}\right)\right\}.
$$
 If $z\to\infty$, then the rate}
$$
R(z,1)\,\ge\,\underline{R}(z,1)=\frac{1}{z^2\log e }(1+o(1))=
\frac{0.693}{z^2}(1+o(1)).
$$

In the first and second rows of Table~2, we give  values
of $\underline{R}(s,1)<1/s$, $s=2,3\dots,8$, along
with the corresponding  values of
$\overline{R}(s,1)<1/s$, $s=2,3\dots,8$,
taken from Table~1.

\smallskip
\begin{center}
\begin{tabular}{||c||c|c|c|c|c|c|c||}
\hline
$s$ & 2 & 3 & 4 & 5 & 6 & 7 & 8\\
\hline
\hline
$\underline{\widetilde{R}}_{1}\,(\le s)=\underline{R}(s,1)$
& .182 & .079 & .044 & .028 & .019 & .014 & .011\\
\hline
$\overline{\widetilde{R}}_{1}\,(\le s)=\overline{R}(s,1)$
& .3219 & .1993 & .1405 & .1056 & .0830 & .0673 & .0559\\
\hline
\hline
$\underline{\widetilde{R}}_{2}\,(\le s)=\underline{R}(s-1,2)$
& - & .0321 &  .0127 & .0068 & .0037 & .0024 & .0015\\
\hline
$\overline{\widetilde{R}}_{2}\,(\le s)=\overline{R}(s-1,2)$
& - & .1610 & .0745 & .0455 & .0287 & .0204 & .0146\\
\hline
\hline
$\underline{\widetilde{R}}_{3}\,(\le s)=\underline{R}(s-2,3)$
& - & - &  .0127 & .0046 & .0020 & .0010 & .0001\\
\hline
$\overline{\widetilde{R}}_{3}\,(\le s)=\overline{R}(s-2,3)$
& - & -  & .0745 & .0387 & .0183 & .0109 & .0067\\
\hline
\hline
$\underline{R}\left(F_0^{1},=s\right)$ & .302 & .142 & .082 & .053 & .037 & .027 &
.021\\
\hline
\end{tabular}
\end{center}
\centerline{\textbf{Table~2}}

\section{$\left(F^{\l},\le s\right)$--Designs,
$\left(F^{\l},=s\right)$--Designs and  $\D_s^{\l}$--Codes}
\quad
In this section we introduce the concept of
non-adaptive group testing designs arising from the potentialities of compressed
genotyping models in  molecular biology and establish a universal
upper bound on their rate. The universal bound is prescribed by
our recurrent upper bound on the rate of classical superimposed
codes. Using notations of Section~1, we give

\textbf{Definition 2.} 
\quad  Let  $\l$, $s$, $t$ be  integers with  $\,1\le \l<s<t$  and
$F^{\l}=F^{\l}(n)$ 
be  an arbitrary fixed function of integer argument $n=0,1,\dots,\l\,$ such that
for any $\,n=0,1,\dots,\l-1$, its value $F^{\l}(n)\ne F^{\l}(\l)$.
Define the  vector
$$
\y^{\l}(P,\A)\,\eq\,
 \left(y^{\l}_1, y^{\l}_2,\dots, y^{\l}_N\right),\quad
 y^{\l}_i\eq\cases{F^{\l}(n) & if $|P\cap A_i|=n$,
\quad $n=0,1,\dots,\l-1$,
\cr
F^{\l}(\l) & if $|P\cap A_i|\ge \l$,\quad $i\in[N]$,\cr}
$$
or
$$
\y^{\l}(P,X)\,\eq\,
 \left(y^{\l}_1, y^{\l}_2,\dots, y^{\l}_N\right),\quad
 y^{\l}_i\eq\cases{F^{\l}(n) & if $\sum\limits_{m=1}^{|P|}\,x_i(p_m)=n$,
\quad $n=0,1,\dots,\l-1$,
\cr
F^{\l}(\l) & if $\sum\limits_{m=1}^{|P|}\,x_i(p_m)\ge \l$,\quad $i\in[N]$.\cr}
$$
A code $X$ of length $N$ and size $t$    
is called an {\em $\left(F^{\l},\le s\right)$--design}, ({\em $\left(F^{\l},=s\right)$--design}),
$1\le \l< s<t$,
 {\em for  group testing model}
if $\y^{\l}(P,X)\,\ne\,\y^{\l}(P',X)$ for any
$$
P\ne P',\quad P\in{\cal P}(t,\le s),\;
P'\in{\cal P}(t,\le s)\quad \left(P\in{\cal P}(t,=s),\;
P'\in{\cal P}(t,=s)\right).
$$

\textbf{Remark 1}.\quad
 $\left(F^{\l},\le s\right)$--design  and $\left(F^{\l},=s\right)$--design
 are  examples,
which  can be interpreted as compressed genotyping~\cite{yaniv10} models in
molecular biology.

\textbf{Remark 2}.\quad
In \cite{EM11}, a special $\left(F^{\l},\le s\right)$--design is considered.
The authors introduce the ranges $(0\eq r_0<r_1<r_2<\dots<r_k\eq p)$ and set
\begin{eqnarray*}
F^\l(r_0+1)  &=& \dots\  = F^\l(r_1)=1\\
F^\l(r_1+1)    &=& \dots\  = F^\l(r_2)=2\\
\vdots         &=& \dots\  = \vdots\\
F^\l(r_{k-1}+1) &=& \dots\  = F^\l(r_k)=k.
\end{eqnarray*}
This model can be viewed as an adder model followed by a
quantizer.
\bigskip

Let  $1\le \l< s<t$ be integers.
For any set $\S\subset[t]$ of size  $|\S|=s$, we denote
by ${\S\choose \l}$ the collection of all ${s\choose \l}$
$\l$--subsets of the set $\S$.
\medskip

\textbf{Definition 3.} \cite{dr84}.  A family of subsets
$\Omega_1,\Omega_2,\dots,\Omega_t$ is called
an $\D_s^{\l}$--{\em  family} if for any  $\S\subset[t]$, $|\S|=s$, and
any $j\not\in\S$, 
$$
\Omega_j \not\subseteq
\bigcup_{{\S\choose \l}}\left\{\bigcap_{k=1}^{\l}\,
\Omega_{j_k}\right\},
\;\mbox{where}\;
{\S\choose \l}\eq\left\{(j_1,j_2,\dots,j_{\l})\;:\;
j_i\in\S,\quad j_1<j_2<\dots<j_{\l}\right\}.
$$
 An incidence matrix
$X=\|x_i(j)\|,\;i\in[N],\;j\in[t]$,
corresponding to  $\D_s^{\l}$-- 
family is called a
{\em superimposed $\D_s^{\l}$-code} (briefly,
{\em  $\D_s^{\l}$-code}).

One can easily check the following

\textbf{Proposition 2.} {\em Any binary $(N\times t)$-matrix
$X$ is a  $\D_s^{\l}$--code, $1\le \l< s<t$, if and only if
for any collection of $s+1$ integers
$j_1,j_2,\dots,j_s,j_{s+1}$, $j_k\ne j_m$,
$j_k\in[t]$, there exists $i\in [N]$ such that
$$
x_i(j_{s+1})=1,\qquad
\sum_{k=1}^s\;x_i(j_k)\le \l-1.
$$
For $\l=1$ and $s=2,3\dots$, the definition of $\D_s^{1}$-code coincides with the definition
of superimposed $(s,1)$-code. In addition, if $1\le\l<s-1$, then
any $\D_s^{\l}$-code is a $\D_s^{\l+1}$-code}.
\medskip

\textbf{Remark 3}.\quad
For $s>\l\ge2$, $\D_s^{\l}$-codes were
suggested in~\cite{dr84} for the study of some
communication systems with random multiple access.

\subsection{Universal Upper Bound for $\left(F^{\l},\le
s\right)$--Designs}
\quad
Let $t\left(N,\D_s^{\l}\right)$,
$t\left(N,F^{\l},\le s\right)$ and
$t\left(N,F^{\l},=s\right)$
 be the maximal  size
 of superimposed  $\D_s^{\l}$--codes,
  $\left(F^{\l},\le s\right)$--designs and
 $\left(F^{\l},=s\right)$--designs.
 For fixed $1\le\l<s$, define the corresponding {\em rates}:
$$
R\left(\D_s^{\l}\right)\eq\varlimsup_{N\to\infty}\,
\frac{\log_2t\left(N,\D_s^{\l}\right)}{N},\quad 1\le \l<s,
$$
$$
R\left(F^{\l},\le s\right)\eq\varlimsup_{N\to\infty}
\frac{\log_2t\left(N,F^{\l},\le s\right)}{N},\qquad
R\left(F^{\l},=s\right)\eq\varlimsup_{N\to\infty}
\frac{\log_2t\left(N,F^{\l},=s\right)}{N}.
$$
Obviously, for any $1\le\l<s$, the following inequalities hold:
$$
t\left(N,F^{\l},\le s\right)\,\le\,t\left(N,F^{\l},=s\right),\quad
R\left(F^{\l},\le s\right)\,\le\,R\left(F^{\l},=s\right)\,\le\,\frac{\log_2(\l+1)}{s}.
\eqno(4)
$$
\smallskip

\textbf{Proposition 3.} \cite{dr84}.
{\em
If $1\le\l<s-1$, then any $\left(F^{\l},\le s\right)$--design is a superimposed
 $\D_{s-1}^{\l}$--code,
i.e.,}
$$
t\left(N,F^{\l},\le s\right)\,\le\, t\left(N,\D_{s-1}^{\l}\right),\qquad
R\left(F^{\l},\le s\right)\,\le\,R\left(\D_{s-1}^{\l}\right),\quad 1\le\l<s-1.
$$

\textbf{Proof}.\quad By contradiction. If a code
$X=\|x_i(j)\|,\;i\in[N],\;j\in[t]$ doesn't satisfy the
definition of  $\D_{s-1}^{\l}$--code, then in virtue of
Proposition~1, there exists a
collection of $s$ integers
$j_1,j_2,\dots,j_{s-1},j_{s}$, $j_k\ne j_m$,
$j_k\in[t]$, such that for any~$i\in [N]$,
$$
x_i(j_{s})=1\quad\Longrightarrow\quad
\sum_{k=1}^{s-1}\;x_i(j_k)\ge \l.
$$
Hence,  for  $(s-1)$-subset
$P\eq\{j_1,j_2,\dots,j_{s-1}\}\subset[t]$ and
$s$-subset
$P'\eq\{j_1,j_2,\dots,j_{s-1},j_{s}\}\subset[t]$,
the  vector $\y^{\l}(P,X)=\y^{\l}(P',X)$.
This contradicts to the  definition of $\left(F^{\l},\le s\right)$--design.
\medskip

\textbf{Theorem 7.} (De Bonis, Vaccaro~\cite{dbv06}).\quad
{\em For any $1\le \l< s$, the rate $R\left(\D_s^{\l}\right)$
of superimposed  $\D_s^{\l}$--codes satisfies inequality
$$
R\left(\D_s^{\l}\right)\,\le\,
R\left(\left\lfloor\frac{s}{\l}\right\rfloor,1\right),
$$
where $R(z,1)$, $z\ge1$, is the rate of classical superimposed
$(z,1)$-codes}.

 Proposition~3  and Theorem~7 lead to inequalities:
$$
R\left(F^{\l},\le s\right)\,\le\,R\left(\D_{s-1}^{\l}\right)
\,\le\,
R\left(\left\lfloor\frac{s-1}{\l}\right\rfloor,1\right)
\,\le\,
\overline{R}\left(\left\lfloor\frac{s-1}{\l}\right\rfloor,1\right),
\quad 1\le \l\le s,\eqno(5)
$$
where $\overline{R}(z,1)$ is the recurrent upper bound on the rate
$R(z,1)$ presented by Theorem~1. For instance, if $(\l=3,s=10)$ or
$(\l=3,s=13)$,  then  Table~2 shows that
$$
\overline{R}(3,1)=.199<.200=2/10\quad\mbox{or}\quad
\overline{R}(4,1)=.140<.154=2/13,
$$
i.e., for $\l=3$ and $s=3k+1$, $k=3,4,\dots$,  bound~(5) improves the trivial bound~(4).

From inequalities (4)-(5), it follows

\textbf{Proposition 4}. \quad (Universal upper bound).
{\em For any $\left(F^{\l},\le s\right)$--design, the rate
$$
R\left(F^{\l},\le s\right)
\,\le\,
\min\left\{\frac{\log_2(\l+1)}{s}\,;\,
\overline{R}\left(\left\lfloor\frac{s-1}{\l}\right\rfloor,1\right)\right\},
\quad 1\le \l< s,
$$
and the asymptotic inequality
$$
R\left(F^{\l},\le s\right)\,\le\,\frac{2\l^2\log_2s}{s^2}
\,(1+o(1)),\quad \l=1,2,\dots,\qquad s\to\infty,
$$
holds}.

\section{Constructions of Superimposed $(z,u)$-Codes and $\D_s^{\l}$--Codes}
\subsection{Superimposed
$(s,1)$-Codes and $\D_s^{\l}$--Codes  Based on\\
Shortened  Reed-Solomon Codes}
\quad
Let ${\cal Q}$ be the set of all primes or
prime powers $\ge2$, i.e.,
$$
{\cal Q}\eq
\{2,3,4,5,7,8,9,11,13,16,17,19,23,25,27,29,31,32,37,\dots\}.
$$
Let $q\in{\cal Q}$ and $2\le k\le q+1$ be fixed
integers for which there exists the $q$-ary
Reed-Solomon code (RS-code)
$B$ of size $q^{k}$, length $(q+1)$ and
the Hamming distance
$
d=q-k+2=(q+1)-(k-1)
$~\cite{ms77}.
We will identify
the code $B$ with an
$\left((q+1)\times q^{k}\right)$--matrix whose
columns, (i.e., $(q+1)$-sequences from the alphabet
$\{0,1,2,\dots,q-1\}$) are the codewords
of $B$.
Therefore, the maximal
possible number of positions (rows) where its two
codewords (columns) can coincide, called a {\em
coincidence} of code $B$, is equal to $k-1$.
\medskip

Fix an arbitrary integer $r=0,1,2,\dots,k-1$
and introduce
the {\it shortened} RS-code
$\tilde B$ of size $t=q^{k-r}$,
length $n=q+1-r$ that
has the same Hamming
distance $d=q-k+2$. Code $\tilde B$
is obtained by the {\it shortening} of the {\it subcode} of
$B$ which contains $0's$ in the first $r$ positions (rows)
of $B$.  Obviously, the coincidence of $\tilde B$
is equal to
$$
\lambda\eq n-d=
(q+1-r)-d=q+1-r-(q-k+2)=k-r-1.
\eqno(6)
$$

Consider the following standard transformation of the
$q$-ary code $\tilde B$, when each symbol of the
$q$-ary alphabet $\{0,1,2,\dots,q-1\}$ is substituted
for the corresponding binary column of the length $q$ and
the weight $1$, namely:
$$
0\Leftrightarrow\underbrace{(1,0,0,\dots,0)}_{q},\quad
1\Leftrightarrow\underbrace{(0,1,0,\dots,0)}_{q},\quad
\dots\quad
q-1\Leftrightarrow\underbrace{(0,0,0,\dots,1)}_{q}.
$$
As a result we have a binary constant-weight
code $X$ of size $t$, length $N$ and weight $w$,
where
$$
t=q^{k-r}=q^{\lambda+1},\quad N=n\cdot q=(q+1-r)q,\quad
w=n=q+1-r. \eqno(7)
$$
From Propositions~1-2 and (6), it follows

\textbf{Proposition 5}.  {\em Let integers  $1\le \l< s$
satisfy inequalities
$$
s[(k-1)-r]\,\le\,\l\,(q+1-r)\,-\,1,\quad 2\le k\le q+1,\quad
0\le r\le k-1. \eqno(8)
$$
Then the binary constant-weight code $X$ with parameters $(7)$
is a $\D_s^{\l}$--code if $\,2\le \l<s,\,$ or $X$ is a superimposed
$(s,1)$-code if~$\l=1$.}
\medskip

For $\l=1$, the detailed tables with parameters of the best known
superimposed $(s,1)$-codes (or $\D_s^{1}$--codes)
based on  Proposition~5
are presented in our papers~\cite{dmr00_1}-\cite{dmr00_2}.
 Table~3 gives an example of such table.
In Table~3, we marked by the {\bf boldface type} two {\em  triples}  of
 superimposed code parameters which were known
from~\cite{ks64}. The rest triples of superimposed code parameters
from  Table~3 were obtained in~\cite{dmr00_1}-\cite{dmr00_2}.
\medskip

For the general case of  superimposed $(z,u)$-codes, $2\le u< z$,
the construction similar to Proposition~5 was developed in~\cite{dmtv02}.
Another significant constructions of  superimposed $(z,u)$-codes, $2\le u< z$,
were suggested in~\cite{KL04}-\cite{lk04}.  Table~4 gives
parameters of the best known
superimposed $(z,u)$-codes if $u=2,3$ and $z=2,3,\dots9$.


\subsection{ Parameters of constant-weight superimposed
$(s,1)$-codes  $\;2\le s\le 8$, of
weight $w$,  length $N$, size $t=q^{\lambda+1}$,
$2^m\le t<2^{m+1}$, $5\le m\le30$, based on the
$q$-ary shortened Reed-Solomon codes.}

\begin{center}
\begin{tabular}{|c||c|c|c|c|c|c|c|}
\hline
$s$ & 2 & 3 & 4 & 5 & 6 & 7 & 8\\
\hline
$\underline{R}(s,1)$  &$.182$
&$.079$ &$.044$
&$.028$ &$.019$
&$.014$ &$.011$\\
\hline
$\overline{R}(s,1)$ &$.322$ &
$.199$ &$.140$ &
$.106$ &$.083$ &
$.067$ &$.056$\\
\hline
$m$ & $q,\,\lambda,\,N$ &
$q,\,\lambda,\,N$ & $q,\,\lambda,\,N$ &
$q,\,\lambda,\,N$ & $q,\,\lambda,\,N$ &
$q,\,\lambda,\,N$ & $q,\,\lambda,\,N$\\
\hline
$5$ & $-$ & $7,\,1,\,28$ & $7,\,1,\,35$
& $7,\,1,\,42$ & $7,\,1,\,49$ & $-$ & $-$ \\
\hline
$6$ & $4,\,2,\,20$ &
$8,\,1,\,32$ & $8,\,1,\,40$ & $8,\,1,\,48$
& $8,\,1,\,56$ & $9,\,1,\,72$ & $11,\,1,\,99$ \\
\hline
$7$ & $-$ &
$-$ & $13,\,1,\,65$ &$13,\,1,\,78$&
$13,\,1,\,91$&$13,\,1,\,104$&$13,\,1,\,117$\\
\hline
$8$ & $7,\,2,\,35$ &
$7,\,2,\,49$ & $-$ &
$16,\,1,\,96$ & $16,\,1,\,112$&$16,\,1,\,128$
&$16,\,1,\,144$\\
\hline
$9$ & $8,\,2,\,40$ &
$8,\,2,\,56$ & $8,\,2,\,72$ &
$-$ & $23,\,1,\,161$ &
$23,\,1,\,184$ & $23,\,1,\,207$\\
\hline
$10$ & $-$ &
$11,\,2,\,77$ & $11,\,2,\,99$ &
$11,\,2,\,121$ & $-$ &
$-$ & $-$\\
\hline
$11$ & $7,\,3,\,49$ &
$-$ & $13,\,2,\,117$ &
$13,\,2,\,143$ & $13,\,2,\,169$ &
$-$ & $-$\\
\hline
$12$ & $8,\,3,\,56$ &
$9,\,3,\,90$ & $16,\,2,\,144$ &
$16,\,2,\,176$ & $16,\,2,\,208$ &
$16,\,2,\,240$ & {\bf16,\,2,\,272}\\
$\frac{12}{N}$ &.214 &.133 &.083 &.068 &.058 &.050 &
.044\\
\hline
$13$ & $-$ &
$11,\,3,\,110$ & $-$ &
$23,\,2,\,253$ & $23,\,2,\,299$ &
$23,\,2,\,345$ & $23,\,2,\,391$\\
\hline
$14$ & $-$ &
$13,\,3,\,130$ & $13,\,3,\,169$ &
$-$ & $27,\,2,\,351$ &
$27,\,2,\,405$ & $27,\,2,\,459$\\
\hline
$15$ & $8,\,4,\,72$ &
$-$ & $-$ &
$-$ & $-$ &
$32,\,2,\,480$ & $32,\,2,\,544$\\
\hline
$16$ & $-$ &
$16,\,3,\,160$ & $16,\,3,\,208$ &
$16,\,3,\,256$ & $19,\,3,\,361$ &
$-$ & $-$\\
\hline
$17$ & $11,\,4,\,99$ &
$-$ & $-$ &
$-$ & $-$ &
$-$ & $-$\\
\hline
$18$ & $13,\,4,\,117$ &
$13,\,4,\,169$ & $-$ &
$23,\,3,\,368$ & $23,\,3,\,437$ &
$23,\,3,\,506$ & $25,\,3,\,625$\\
\hline
$19$ & $-$ &
$-$ & $-$ &
$27,\,3,\,432$ & $27,\,3,\,513$ &
$27,\,3,\,594$ & $27,\,3,\,675$\\
\hline
$20$ & $11,\,5,\,121$ &
$16,\,4,\,208$ & {\bf16,\,4,\,272} &
$-$ & $32,\,3,\,608$ &
$32,\,3,\,704$ & $32,\,3,\,800$\\
$\frac{20}{N}$ &.165 &.096 &.074 &-&.034 &.028 &
.025\\
\hline
$21$ & $-$ &
$-$ & $19,\,4,\,323$ &
$-$ & $-$ &
$-$ & $41,\,3,\,1025$\\
\hline
$22$ & $13,\,5,\,143$ &
$-$ & $23,\,4,\,391$ &
$23,\,4,\,483$ & $-$ &
$-$ & $-$\\
\hline
$23$ & $-$ &
$-$ & $25,\,4,\,425$ &
$25,\,4,\,525$ & $25,\,4,\,625$ &
$-$ & $-$\\
\hline
$24$ & $-$ &
$16,\,5,\,256$ & $-$ &
$27,\,4,\,609$ & $29,\,4,\,725$ &
$29,\,4,\,841$ & $-$\\
\hline
$25$ & $13,\,6,\,169$ &
$19,\,5,\,304$ & $-$ &
$-$ & $32,\,4,\,800$ &
$32,\,4,\,928$ & $32,\,4,\,1056$\\
$\frac{25}{N}$ &.148 &.082 &-&-&.031 &.027 &
.024\\
\hline
$26$ & $-$ &
$-$ & $-$ &
$-$ & $37,\,4,\,925$ &
$37,\,4,\,1073$ & $37,\,4,\,1221$\\
\hline
$27$ & $-$ &
$-$ & $23,\,5,\,483$ &
$-$ & $-$ &
$43,\,4,\,1247$ & $43,\,4,\,1419$\\
\hline
$28$ & $16,\,6,\,208$ &
$-$ & $27,\,5,\,702$ &
$25,\,5,\,650$ & $-$ &
$-$ & $49,\,4,\,1617$\\
\hline
$29$ & $-$ &
$19,\,6,\,361$ & $29,\,5,\,609$ &
$29,\,5,\,754$ & $31,\,5,\,961$ &
$-$ & $-$\\
$\frac{29}{N}$ &$-$&.080 &.048 & .038 & .030 &$-$ &
$-$\\
\hline
$30$ & $-$ &
$-$ & $-$ &
$32,\,5,\,832$ & $32,\,5,\,992$ &
$-$ & $-$\\
\hline
\hline
\end{tabular}
\end{center}
\centerline{\textbf{Table~3}}

Table~3 also contains numerical values of the rate
for several obtained codes, namely: the values of
fraction $\frac{m}{N},\;m=12,\,20,\,25,\,29$.
The comparison with lower $\underline{R}(s,1)$ and
upper $\overline{R}(s,1)$  bounds
from Table~2 (their  values are included in
Table~3 as well) yields the following conclusions:
\begin{itemize}
\item
if $s=2$ and $m\le15$, then the values $\frac{m}{N}$
exceed the random coding rate
$\underline{R}(2,1)=.182$;
\item
if $s\ge3$ and $m\le30$, then the values $\frac{m}{N}$
exceed the random coding rate
$\underline{R}(s,1)$.
\end{itemize}

\subsection{Size $t$ and Length $N$ of  Superimposed
$(z,u)$-Codes, $u=2,3$ and $z=2,3,\dots9$}

\begin{center}
\begin{tabular}{|c||c|c|c|c|c|c|c|}
\hline
(2,2) & (3,2) & (4,2) & (5,2) & (6,2) & (7,2) & (8,2) & (9,2)\\
\hline $t$, $N$ & $t$, $N$ & $t$, $N$ & $t$, $N$ & $t$, $N$ & $t$,
$N$ & $t$, $N$ & $t$, $N$\\
\hline 8, 14 & 7, 21 & 11, 55 & 11, 55
& 20, 190 & 26, 260 & 16, 120 & 38, 703 \\
\hline 9, 18 & 8, 28 & 13, 65 & 16, 120 & 25, 210
& 50, 350 & 32, 496 & 82, 738 \\
\hline 10, 20 & 10, 30 & 17, 68 & 26, 130 & 49, 294 &
64, 448 & 65, 520 & 120, 1090 \\
\hline 12, 22 & 16, 42 & 22, 77 & 48, 246 & 63, 385 &
80, 568 & 81, 648 & 166, 1562 \\
\hline 16, 26 & 21, 56 & 25, 100 & 62, 330 & 79, 497 &
118, 882 & 119, 981 & 250, 2531 \\
\hline 18, 30 & 24, 76 & 47, 205 & 78, 434 &
117, 792 & 164, 1308 & 165, 1430 & 282, 2933 \\
\hline 22, 34 & 49, 147 & 64, 252 & 121, 605 & 169, 1014
& 256, 1800 & 256, 2040 & 361, 3249 \\
\hline 24, 37 & $-$ & $-$ & $-$ & $-$ & $-$ &
$-$ & $-$\\
\hline 32, 43 & (3,3) & (4,3) & (5,3) & (6,3) & (7,3) & (8,3) & (9,3)\\
\hline 40, 50 & $t$, $N$ & $t$, $N$ & $t$, $N$ & $t$, $N$ & $t$,
$N$ & $t$, $N$ & $t$, $N$\\
\hline 48, 59 & 7, 35 & 12, 220 & 16, 560 & 17, 680 & 19, 969 &
20, 1140 & 22, 1540 \\
\hline 56, 65 & 8, 54 & 13, 253 & 19, 612 &
20, 816 & 21, 1071 & 21, 1330 & 23, 1771 \\
\hline 64, 68 & 11, 66 & 23, 253 & 25, 700 &
26, 910 & 27, 1170 & 22, 1386 & 45, 14190 \\
\hline 80, 76 & 16, 112 & 24, 532 & 31, 3951 & 32, 4683 & 52, 11313
& 53, 12757 & 54, 14352 \\
\hline 112, 96 & 22, 176 & 169, 3289 & 50, 8830 & 51, 10008 & 529,
25740 & 729, 73125 & 729, 81900 \\
\hline 128, 100 & 23, 399 & $-$ & 256, 8960 &
361, 15504 & $-$ & $-$ & $-$\\
\hline 144, 109 & 121, 660 & $-$ & $-$ & $-$ &
$-$ & $-$ & $-$\\
\hline 512, 126 & $-$ & $-$ & $-$ & $-$ &
$-$ & $-$ & $-$\\
\hline
\hline
\end{tabular}
\end{center}
\centerline{\textbf{Table~4}}

\subsection{Examples of $\D_s^{\l}$--Codes}
\quad
\textbf{Example 1.}  If $q=5$, then for the pair
$(\l=2,\,s=3)$, inequalities (8) are fulfilled
at $k=5$ and $r=2$. Therefore, the construction of
Proposition~4 yields  a binary constant-weight
$\D_3^{2}$--code $X$  with parameters
$$
t=q^{k-r}=5^3=125,\quad N=n\cdot q=(q+1-r)q=4\cdot5=20,
\quad
w=n=q+1-r=4. \eqno(9)
$$
Parameters (9) give the following lower bound on the
maximal size:~$t\left(20,\D_3^{2}\right)\ge 125$.

\textbf{Example 2.}  If $q=7$, then for the pair
$(\l=2,\,s=4)$, inequalities (8) are fulfilled
at $k=6$ and $r=3$. Therefore, the construction of
Proposition~4 yields  a binary constant-weight
$\D_4^{2}$--code $X$  with parameters
$$
t=q^{k-r}=7^3=343,\quad N=n\cdot q=(q+1-r)q=5\cdot7=35,
\quad
w=n=q+1-r=5. \eqno(10)
$$
Parameters~(10) give the following lower bound on the
maximal size:~$t\left(35,\D_4^{2}\right)\ge 343$.

\textbf{Example 3.}  If $q=8$, then for two pairs of integers
$(\l=2,\,s=6)$ and $(\l=3,\,s=10)$, inequalities (8) are fulfilled
at $k=5$ and $r=2$. Therefore, the construction of
Proposition~4 yields  a binary constant-weight
$\D_6^{2}$--code $X$
and a binary constant-weight  $\D_{10}^{3}$--code $X$ with parameters
$$
t=q^{k-r}=8^3=512,\quad N=n\cdot q=(q+1-r)q=7\cdot8=56,
\quad
w=n=q+1-r=7. \eqno(11)
$$
Parameters (11) give the following lower bounds on the maximal size
$t\left(N,\D_s^{\l}\right)$ of $\D_s^{\l}$--codes:
$$
t\left(56,\D_6^{2}\right)\ge 512,\qquad t\,\left(56,\D_{10}^{3}\right)\ge 512.
$$
For comparison, if $(u=1,\,z=6)$ and $N=56$, then the best known lower bound
on the size of optimal superimposed $(6,1)$-codes, calculated in~\cite{dmr00_1},
is~$\,t(56,6,1)\ge64$.
In addition, this  example shows that for $\l=3$, the parameter
$s=10$ of $\D_{10}^{3}$--code $X$ can exceed the corresponding code
weight~$w=7$.

\section{Threshold Group Testing Model}
\subsection{Superimposed $(z,u)$-Codes
and $\left(F_0^{\u},\le s\right)$-Designs}
\quad
Let the  function $F^{\u}\eq F_0^{\u}=F_0^{\u}(n)$, $1\le \u<s$, takes binary values, namely:
$$
F_0^{\u}(n)\eq\cases{0 & if
\quad $n=0,1,\dots,\u-1$,
\cr
1 & if \quad $n=\u$.\cr}
$$
If $\u\ge2$, then the given particular case is called a {\em threshold group testing
model}~\cite{dam06}. For the non-adaptive threshold group testing model which is
the principal model  for
applications~\cite{yaniv10},
a {\em refined form} of Definition~2 can
be written as follows.
\bigskip

\textbf{Definition 4.} 
\quad  Let $\u,\,1\le \u<s<t$ be  integers.
For code $X=\|x_i(k)\|$, $k\in[t]$, $i\in[N]$, and a subset $P\in{\cal P}(t,\le s)$,
define the {\em $i$-th outcome of
non-adaptive threshold group testing}
$$
y^{\u}_i\,(P,X)\,\eq\,
 \cases{0 & if $\sum\limits_{k\in P}\,x_i(k)\le\u-1$,
\cr
1 & if $\sum\limits_{k\in P}\,x_i(k)\ge \u$,\quad $i\in[N]$.\cr}
$$
A code $X$ is called a {\em  $\left(F_0^{\u},\le s\right)$-design},
({\em $\left(F_0^{\u},=s\right)$-design}) 
if for any  $P,P'\in{\cal P}(t,\le s)$, $P\ne P'$, and such that
$P,P'\in{\cal P}(t,\le s)\setminus{\cal P}(t,\le\u-1)$
$\left(P\in{\cal P}(t,=s),\;P'\in{\cal P}(t,=s)\right)$,
there exists an index $i\in[N]$, where~$y^{\u}_i\,(P,X)\ne y^{\u}_i\,(P',X)$.
\medskip

An important connection between
$\left(F_0^{\u},\le s\right)$-designs and superimposed
$(s-\u+1,\u)$--codes is described by
\smallskip

\textbf{Proposition 6}. (\cite{chen-fu09},~\cite{leb10}).\quad
{\em If $\;1\le \u< s$, then any superimposed $(s-\u+1,\u)$--code is a
$\left(F_0^{\u},\le s\right)$-design, i.e.}
$$
t\,(N,s-\u+1,\u)\,\le\,
t\left(N,F_0^{\u},\le s\right),\qquad
R(s-\u+1,\u)\,\le\,R\left(F_0^{\u},\le s\right).
$$

The lower bound of Theorem~5 and  Propositions~6 lead to the following
lower bound
on the rate of $\left(F_0^{\u},\le s\right)$--designs.
\smallskip

\textbf{Proposition 7}.\quad
(Random coding bound). {\em For any $1\le\u<s$, the rate}
$$
R\left(F_0^{\u},\le s\right)\,\ge\, R(s-\u+1,\u)\ge -\frac{1}{s}\,
\log_2\left[1-\frac{(s-\u+1)^{s-\u+1}\cdot\u^{\u}}{(s+1)^{s+1}}\right],\quad
1\le\u<s.
\eqno(12)
$$
{\em If $\u\ge1$ is fixed and $s\to\infty$, then the asymptotic form of
the given lower bound is}
$$
R\left(F_0^{\u},\le s\right)\,\ge\frac{e^{-\u}\cdot\u^{\u}\cdot\log_2e}{s^{\u+1}}
\,\cdot(1+o(1)).
\eqno(13)
$$

\subsection{Bounds on the Rate of $\left(F_0^{1},\le s\right)$ and
$\left(F_0^{1},=s\right)$-Designs}
\quad
If $\u=1$ and $s\ge2$, then the  the universal upper bound of
Proposition~4 lead to inequalities :
$$
R\left(F_0^{1},\le s\right)\,\le\,
\min\{1/s\,;\,\overline{R}(s-1,1)\},\qquad s=2,3,\dots,
$$
where $\overline{R}(z,1)$, $z=1,2,\dots$, is the recurrent upper
bound from Theorem~1. Hence,  the asymptotic upper bound
$$
R\left(F_0^{1},\le s\right)\le\overline{R}(s-1,1)=
\frac{2\cdot\log_2s}{s^2}\,\cdot(1+o(1)),\quad s\to\infty,
$$
holds.

In~\cite{drr89}-\cite{dr89} (see, also~\cite{d04}),
we obtained the best known asymptotic  random coding
lower  bounds on  $R\left(F_0^{1},\le s\right)$  and $\,R\left(F_0^{1},=s\right)$
along with the best known  upper bound on~$\,R\left(F_0^{1},=s\right)$.
These bounds have the form:
$$
R\left(F_0^{1},\le s\right)\,\ge\,\underline{R}(s,1)\,=
\,\frac{1}{s^{2}\cdot\log_2e}
\,\cdot(1+o(1))=\frac{0.693}{s^2}\,\cdot(1+o(1)),\quad s\to\infty,
\eqno(14)
$$
$$
R\left(F_0^{1},=s\right)\ge \underline{R}\left(F_0^{1},=s\right)=
\frac{2}{s^{2}\cdot\log_2e}
\,\cdot(1+o(1))=\frac{1.386}{s^2}\,\cdot(1+o(1)),\quad s\to\infty,
\eqno(15)
$$
$$
R\left(F_0^{1},=s\right)\le \overline{R}\left(F_0^{1},=s\right)=
\frac{4\cdot\log_2s}{s^2}\,\cdot(1+o(1)),\quad s\to\infty.
\eqno(16)
$$
Lower bound (14), i.e., function  $\underline{R}(s,1)$  is
defined in Theorem~6. For the particular case $\u=1$,
bound (14) is better than the lower bound~(13) of Proposition~7.
The numerical values of lower bound (15),
i.e., numbers $\underline{R}\left(F_0^{1},=s\right)$, $s=2,3,\dots,8$,
are given in Table~2.
\medskip

In addition, applying the corresponding  non-asymptotic
results~\cite{d04}, one can calculate numerical values of
upper bound (16), i.e., numbers  $\overline{R}\left(F_0^{1},=s\right)$, $s\ge1$,
which lead to inequalities:
$R\left(F_0^{1},=s\right)<1/s\,$ if~$\,s\ge11$.
For $s=2$, the nontrivial inequality
$R\left(F_0^{1},=2\right)<0.4998<1/2\,$ was proved
in~\cite{cop98}. For $3\le s\le10$, the inequality  $R\left(F_0^{1},=s\right)<1/s\,$
 can be considered as our conjecture.
\medskip

\subsection{Lower Bound on the Rate of $\left(F_0^{\u},\le s\right)$--Designs}
\quad
For  $\left(F_0^{\u},\le s\right)$--designs, $\u\ge2$,
the lower bound~(12) of Proposition~7 can be
improved~\cite{mahdi-11}. An improvement is obtained with the
help of the following  auxiliary  concepts.
\smallskip

\textbf{Definition 5.}~\cite{mahdi-11}.
\quad  Let $\u,\,1\le \u<s<t/2$ be  integers.
For code $X=\|x_i(k)\|$, $k\in[t]$, $i\in[N]$, and a subset $P\in{\cal P}(t,\le s)$,
define the {\em $i$-th outcome of
non-adaptive threshold group testing}
$$
y^{\u}_i\,(P,X)\,\eq\,
 \cases{0 & if $\sum\limits_{k\in P}\,x_i(k)\le\u-1$,
\cr
1 & if $\sum\limits_{k\in P}\,x_i(k)\ge \u$,\quad $i\in[N]$.\cr}
$$
A code $X$ is called a {\em  threshold $(\u,\le s)$--design}
  of length $N$ and size $\,t\,$
if for any $P,P'\in{\cal P}(t,\le s)$, $P\ne P'$, and such that
$$
|P|\ge|P'|\quad
P,P'\in{\cal P}(t,\le s)\setminus{\cal P}(t,\le\u-1)
$$
there exists an index $i\in[N]$, where the $i$-th outcome
of non-adaptive threshold group testing is
$$
y^{\u}_i\,(P,X)=1\quad\mbox{and}\quad y^{\u}_i\,(P',X)=0.
$$


Let $\,t_{\u}(N,\le s),\,$  denote  the maximal possible size
of  threshold $\,(\u,\le s)$--designs.
 For fixed  $1\le\u<s$, define the corresponding {\em rate}:
$$
R_{\u}\,(\le s)\eq\varlimsup_{N\to\infty}
\frac{\log_2t_{\u}(N,\le s)}{N}.
$$
Obviously, any threshold $\,(\u,\le s)$--designs is a
$\left(F_0^{\u},\le s\right)$--design and the rate
$$
R\,\left(F_0^{\u},\le s\right)\,\ge\,R_{\u}\,(\le s),\quad
1\le\u<s.
\eqno(17)
$$

\textbf{Definition 6.} \cite{mahdi-11}.
\quad  Let $\u,\,1\le \u<s<t/2$ be  integers. A binary $(N\times t)$-matrix
$X$ is called a {\em superimposed  $\M_s^{\u}$-code} (briefly, {\em $\M_s^{\u}$-code}) if
for any two non-intersecting  subsets $Z,\,U\in{\cal P}(t,\le s)$,
$\,Z\cap U=\emptyset$,
such that $\,\u\le|U|\le s$, $|Z|\le|U|$  and for any element $\,j\in U$,
the matrix $X$ contains a row $\,\x_i=(\,x_i(1),x_i(2)\dots,x_i(t)\,)$,
$\,i\in[N]$, for which
$$
x_i(j)=1,\quad \sum\limits_{k\in U}\,x_i(k)\,=\,\u \quad \mbox{and}\quad
x_i(k)=0\quad\mbox{for all}\quad k\in Z.
$$


Let $\,t\left(N,\M_s^{\u}\right)\,$  denote  the maximal size
of   $\,\M_s^{\u}$--codes.
For fixed  $1\le\u<s$, introduce
$$
R\left(\M_s^{\u}\right)\eq\varlimsup_{N\to\infty}
\frac{\log_2t\left(N,\M_s^{\u}\right)}{N},\qquad 1\le\u<s.
$$
called a {\em rate} of  $\,\M_s^{\u}$--codes.
The evident connection between $\,\M_s^{\u}$--codes and
superimposed $(2s-\u,1)$-codes is given by
\smallskip

\textbf{Proposition 8.} \cite{mahdi-11}.
\quad {\bf1.} {\em Let $2\le s<t/2$. If $\u=1$, then any
$\,\M_s^{1}$--code $X$ of size $t$ is a  superimposed
$(2s-1,1)$-code and, vice versa, any superimposed  $(2s-1,1)$-code
$X$ of size $t$ is a $\,\M_s^{1}$--code,
i.e.,
the rate~$R\left(\M_s^{1}\right)=R(2s-1,1)$.}
\quad {\bf2.}
{\em  If $\,2\le\u< s<t/2$, then any $\M_s^{\u}$-code $X$ of size $t$  is a  superimposed
$(2s-\u,1)$-code, i.e., the
rate~$R\left(\M_s^{\u}\right)\,\le\,R(2s-\u,1)$}.
\medskip

{\textbf{Proposition 9.} \cite{mahdi-11}.
\quad
 {\em If $1\le\u< s<t/2$, then any $\M_s^{\u}$-code $X$ of size $t$ is a
 threshold $\,(\u,\le s)$--design, i.e. the
rate~$R\left(\M_s^{\u}\right)\le R_{\u}\,(\le s)$}.
\medskip

\textbf{Proof of Proposition 9.}
Let $X=\|x_i(k)\|$, $k\in[t]$, $i\in[N]$, be an arbitrary
$\M_s^{\u}$-code. Consider arbitrary  subsets:
$P,P'\in{\cal P}(t,\le s)$, $P\ne P'$, and such that
$$
|P|\ge|P'|,\quad
P,P'\in{\cal P}(t,\le s)\setminus{\cal P}(t,\le\u-1),
\quad \u\le |P|\le s,\quad \u\le |P'|\le|P|.
$$
Fix an arbitrary $j\in P\setminus P'$, $j\notin P'$ and define non-intersecting  subsets
$U\eq P$ and $Z\eq P'\setminus P$. We have
$$
\u\le |U|\le s,\quad j\in U,\quad U\cap Z=\emptyset,
\quad Z\subset P',\quad P'\setminus Z\subset U, \quad |Z|\le |P'|\le|P|=|U|.
$$
Definition~6 of $\M_s^{\u}$-code
implies  that there exists an index $i\in[N]$ such that
$$
\left(\,\sum\limits_{k\in U} x_i(k)=\u,\; \sum\limits_{k\in Z} x_i(k)=0,\;
x_i(j)=1,\;\sum\limits_{k\in P'\setminus Z} x_i(k)\le\u-1\, \right)\;\Rightarrow\;
$$
$$
\;\Rightarrow\;\left(\,\sum\limits_{k\in P} x_i(k)=\u,\;
\sum\limits_{k\in P'} x_i(k)\le\u-1\,\right)\;\Rightarrow\;
\left(\,y_i(P,X)=1,\; y_i(P',X)=0\,\right),
$$
i.e., code $X$ is a threshold $\,(\u,\le s)$--design.

\qquad Proposition~9 is proved.
\bigskip

If $\beta\eq\Pr\{x_i(k)=1\}$ and
 $1-\beta\eq\Pr\{x_i(k)=0\}$, then one can easily check
that for any $j\in[t]$, the probability
$$
\Pr\left\{\x(j)\; \mbox{is}\; \M_s^{\u}-\mbox{bad}\;
\right\}\,\le\,\sum\limits_{u=\u}^s\,\sum\limits_{z=0}^u\,
{t-1\choose u+z-1}\,{u+z-1\choose u-1}\,\times
$$
$$
\times\,\left[1-{u-1\choose
\u-1}\,\beta^{\u}\,(1-\beta)^{u+z-\u}\right]^N.
$$
The given inequality leads to the following {\em random coding lower bound} on the rate
of $\M_s^{\u}$-codes:

\textbf{Proposition 10.}\quad {\em For any $\beta$, $0<\beta<1$, the rate
$R\left(\M_s^{\u}\right)$ satisfies inequality
$$
R\left(\M_s^{\u}\right)\,\ge\,
\min\limits_{\u\le u\le s;\; 0\le z\le u}\,
\left\{\frac{-\log_2
\left[1-
\,{u-1\choose
\u-1}\,\beta^{\u}\,(1-\beta)^{u+z-\u}
\right]}
{u+z-1}\right\}\ge
\min\limits_{\u\le u\le s}\,L_{\u}(\beta,u),
$$
where}
$$
L_{\u}(\beta,u)\,\eq\,\left\{\frac{-\log_2
\left[1-\,{u-1\choose
\u-1}\,\beta^{\u}\,(1-\beta)^{2u-\u}
\right]}
{2u-1}\right\},\quad \u\le u\le s,\quad 0<\beta<1.\eqno(18)
$$

From~(17) and Propositions~9-10 it follows a lower bound
on the  rate of $\left(F_0^{\u},\le s\right)$-designs :
$$
R\left(F_0^{\u},\le s\right)\,\ge\,
\underline{R}\left(F_0^{\u},\le s\right)\eq
\max\limits_{0<\beta<1}\,
\min\limits_{\u\le u\le s}\,L_{\u}(\beta,u)=
$$
$$
=\max\limits_{0<\beta<1}\,
\min\limits_{\u\le u\le s}\,
\left\{\frac{-\log_2
\left[1-
\,{u-1\choose\u-1}\,\beta^{\u}\,(1-\beta)^{2u-\u}
\right]}
{2u-1}\right\}, \quad 1\le\u<s.
\eqno(19)
$$

The calculation of numerical values for lower bound (19)
is an  open problem.

\subsection{Comments on Definitions 4 and 5}
\quad
Let $\u$, $1\le \u<s<t/2$, be integers.
For  a comparison of Definitions~4 and~5 , introduce

\textbf{Definition $\widetilde{{\bf5}}$.}
\quad
A code $X$ is called a {\em  threshold $\overline{(\u,\le s)}$--design},
  of length $N$ and size $\,t\,$
if for any $P,P'\in{\cal P}(t,\le s)$, $P\ne P'$, and such that
$$
P\setminus P'\ne\emptyset,\quad
P,P'\in{\cal P}(t,\le s)\setminus{\cal P}(t,\le\u-1),
\qquad
$$
there exists an index $i\in[N]$, where the $i$-th outcome
of non-adaptive threshold group testing is
$$
y^{\u}_i\,(P,X)=1\quad\mbox{and}\quad y^{\u}_i\,(P',X)=0.
$$
\medskip

Let $\,\widetilde{t}_{\u}(N,\le s),\,$ be the maximal  size
of  threshold $\,\overline{(\u,\le s)}$--designs.
 For fixed  $1\le\u<s$, define the corresponding {\em rate}
$$
\widetilde{R}_{\u}\,(\le s)\eq\varlimsup_{N\to\infty}
\frac{\log_2\widetilde{t}_{\u}(N,\le s)}{N}.
$$
The following important property is given by
\smallskip

\textbf{Proposition 11.}
\quad {\em If $1\le\u< s<t/2$, then {\bf(1)} any superimposed
$\,(s-\u+1,\u)$--code $X$ of size~$t$ is a
threshold $\,\overline{(\u,\le s)}$--design
and, vice versa, {\bf(2)} any threshold $\,\overline{(\u,\le s)}$--design
$X$ of size~$t$ is a superimposed $\,(s-\u+1,\u)$--code, i.e.,
the rate~$\widetilde{R}_{\u}\,(\le s)\,=\,R(s-\u+1,\u)$.}
\medskip

Evidently,  any threshold $\,\overline{(\u,\le s)}$--design is
a threshold $\,(\u,\le s)$--design. Therefore, in virtue of Proposition~11, the  rate
$$
\widetilde{R}_{\u}\,(\le s)\,=\,R(s-\u+1,\u)\,\le\, R_{\u}\,(\le s)\,
\le\,R\left(F_0^{\u},\le s\right).
$$

Denote by $\underline{R}(z,u)$, $1\le u\le z$,  the lower bound on
$R(z,u)$ formulated in Theorems~5 and~6.
Let $\overline{R}(z,u)$ be the upper bound on
$R(z,u)$ given by Theorem~3.
For parameters $\u=1,2,3$ and $s=\u+1,\u+2,\dots,8$, numerical values of
lower bound
$\underline{\widetilde{R}}_{\u}\,(\le s)\eq\underline{R}(s-\u+1,\u)$
and upper bound
$\overline{\widetilde{R}}_{\u}\,(\le s)\eq\overline{R}(s-\u+1,\u)$
on the rate~$\widetilde{R}_{\u}\,(\le s)\,=\,R(s-\u+1,\u)$
are presented in  Table~2.
\medskip

\textbf{Proof of Proposition 11.}\quad
{\bf(1)}$\,$ Let $X=\|x_i(k)\|$, $k\in[t]$, $i\in[N]$, be a
superimposed $\,(s-\u+1,\u)$--code. Consider arbitrary  subsets:
$P,P'\in{\cal P}(t,\le s)$, $P\ne P'$, and such that
$$
P\setminus P'\ne\emptyset,\quad
P,P'\in{\cal P}(t,\le s)\setminus{\cal P}(t,\le\u-1),
\quad \u\le |P|\le s,\quad \u\le |P'|\le s.
$$
Fix an arbitrary subset $U\subset P$ such that $|U|=\u$, and
$U\setminus P'\ne\emptyset$. Note that the size of intersection
$|P'\cap U|\le \u-1$.

Consider the set $P'\setminus(P'\cap U)$.
Introduce a set $Z$, $Z\subset[t]$, of size $|Z|=s-(\u-1)$,
where the intersection $Z\cap U=\emptyset$,
as follows.
\begin{enumerate}
\item
If  $|P'\setminus(P'\cap U)|\ge s-(\u-1)$, then we choose  the set $Z$,
$Z\subseteq P'\setminus(P'\cap U)$, $Z\cap U=\emptyset$, as an
arbitrary fixed subset of size~$|Z|=s-(\u-1)$. Let a row $i$,
$i\in[N]$ corresponds to the pair $(U,Z)$ in Definition~1 of
superimposed $\,(s-\u+1,\u)$--code~$X$. One can easily see that
$$
\sum\limits_{k\in P} x_i(k)\ge
\,\sum\limits_{k\in U} x_i(k)=\u,\;
\sum\limits_{k\in P'} x_i(k)\,\le\,|P'|-|Z|\le s-[s-(\u-1)]=\u-1.
$$
Hence, $\left(\,y_i(P,X)=1,\; y_i(P',X)=0\,\right)$.

\item
If  $|P'\setminus(P'\cap U)|< s-(\u-1)$, then we choose  the set $Z$,
$Z\supset P'\setminus(P'\cap U)$, as an
arbitrary fixed superset of size~$|Z|=s-(\u-1)$.
 Let a row $i$,
$i\in[N]$ corresponds to the pair $(U,Z)$ in Definition~1 of
superimposed $\,(s-\u+1,\u)$--code~$X$. One can easily see that
$$
\sum\limits_{k\in P} x_i(k)\ge
\,\sum\limits_{k\in U} x_i(k)=\u,\;
\sum\limits_{k\in P'} x_i(k)\,=\,|P'\cap U|\le \u-1.
$$
Hence,  $\left(\,y_i(P,X)=1,\; y_i(P',X)=0\,\right)$.
\end{enumerate}
Arguments~1. and~2. imply that  code $X$ is a threshold $\,\overline{(\u,\le s)}$--design.
Therefore, the statement~{\bf(1)} of Proposition~11 is proved.

{\bf(2)}\quad Let $X=\|x_i(k)\|$, $k\in[t]$, $i\in[N]$, be a
threshold $\,\overline{(\u,\le s)}$--design.
Consider two arbitrary non-intersecting  sets $U$ and $Z$, where
$$
U\subset [t],\quad |U|=\u, \quad
Z\subset [t],\quad |Z|=s-(\u-1),\quad
U\cap Z=\emptyset,
$$
and fix an element $j\in U$. Introduce subsets
$P,P'\in{\cal P}(t,\le s)\setminus{\cal P}(t,\le \u-1)$
as follows:
$$
P\eq U,\quad P'\eq (U\setminus j)\cup Z,\quad
P\setminus P'\ne\emptyset,\quad |P|=\u,\quad
|P'|=(\u-1)+s-(\u-1)=s.
$$
Definition~$\widetilde{{\bf5}}$ of threshold $\,\overline{(\u,\le s)}$--design
means that there exists an index $i\in[N]$ such that
$$
\left(\,y_i(P,X)=1,\; y_i(P',X)=0\,\right)\;\Rightarrow\;
\left(\,\sum\limits_{k\in P} x_i(k)\ge\u,\;\sum\limits_{k\in P'}
x_i(k)\le\u-1\right)\;\Rightarrow\;
$$
$$
\;\Rightarrow\;\left(\,\sum\limits_{k\in U} x_i(k)\ge\u,\;
\sum\limits_{k\in U\setminus j} x_i(k)+
\sum\limits_{k\in Z} x_i(k)\le \u-1\right)\;\Rightarrow\;
$$
$$
\;\Rightarrow\;\quad x_i(k)=1,\; k\in U,\;|U|=\u;
\qquad x_i(k)=0,\; k\in Z,\; |Z|=s-(\u-1).
$$
Hence, code $X$ is a superimposed  $(s-\u+1,\u)$-code, i.e.,
statement {\bf(2)} is established.

\qquad Proposition~11 is proved.

\section{Concluding Remarks}
\quad

In this Section, we would like to distinguish the
 principal  achievements for the theory of non-adaptive
group testing models and  superimposed codes obtained in the last
 decade.
\begin{enumerate}
\item
In 2003, Vladimir Lebedev~\cite{l03} proved Theorem~3 which
established a  recurrent inequality for the rate $R(z,u)$ of
superimposed $(z,u)$-codes. This  inequality and the best known numerical values~\cite{dr82,dmtv02} of
upper bound on the rate $R(z,1)$ gave  the best known numerical values of
upper bound on the rate $R(z,u)$, $z\ge\,u\,\ge2$.

\item
In 2004, Vladimir Lebedev and Hyun Kim~\cite{lk04} presented the
best known and optimal constructions (see, Table~4) of superimposed
$(z,u)$-codes,~$z\ge\,u\,\ge2$.

\item
In 2004, Annalisa De Bonis and Ugo Vaccaro~\cite{dbv06} proved
Theorem~7 which established
an  upper bound on the rate of superimposed $\D_s^{\u}$-codes
via the rate $R(z,1)$ of superimposed $(z,1)$-codes. The result
leads to the universal upper bound (Proposition~4) on the rate of group testing
designs motivated by compressed
genotyping models in  molecular biology.

\item
In 2010, Mahdi Cheraghchi~\cite{mahdi-11}
introduced the  concepts of threshold $\,(\u,\le s)$--designs and
superimposed $\,\M_s^{\u}$-codes
and proved Proposition~9 which actually  established an improved lower bound
(19) on the rate of non-adaptive threshold group testing model.
\end{enumerate}

\section*{Acknowledgement}

All authors are grateful to Professor Rudolf Ahlswede for his lifetime
friendship and encouragement. He wrote in 1979 the first book
on search theory in German (\cite{such}), describing
the connection between several areas.
The extensive literature is presented in such a
way that the reader can quickly understand the range of questions and
obtain a survey of them which is as comprehensive as possible.
In 1982 the Russian
edition was published by MIR. It includes also a supplement,
{\it Information-theory Methods in Search Problems}, which was written
by Maljutov.
The English edition, published in 1987, includes also a section
``Further reading'',
where articles and books are mentioned
which inform the researcher about new developments and results which
seem to carry the seed for further discoveries.


\end{document}